\documentclass[prl,aps,twocolumn,floatfix,showpacs]{revtex4}
\usepackage{graphicx,graphics,psfrag,amsmath,calc}
\usepackage{epsfig}
\usepackage{color}
\begin{document}
\title{
Topological phase transitions in ultra-cold Fermi superfluids: \\
the evolution from BCS to BEC under artificial spin-orbit fields}

\author{Kangjun Seo, Li Han and C. A. R. S{\'a} de Melo}
\affiliation{School of Physics, Georgia Institute of Technology,
Atlanta, Georgia 30332, USA}
\date{\today}

\begin{abstract}
We discuss topological phase transitions in ultra-cold Fermi 
superfluids induced by interactions and artificial spin orbit
fields. We construct the phase diagram for population imbalanced 
systems at zero and finite temperatures, and analyze spectroscopic 
and thermodynamic properties to characterize various phase transitions. 
For balanced systems, the evolution from BCS to BEC superfluids in 
the presence of spin-orbit effects is only a crossover as the system remains
fully gapped, even though a triplet component of the order 
parameter emerges.
However, for imbalanced populations, spin-orbit fields induce 
a triplet component in the order parameter that produces nodes 
in the quasiparticle excitation spectrum leading 
to bulk topological phase transitions of the Lifshitz type. 
Additionally a fully gapped phase exists, where a crossover from
indirect to direct gap occurs, but a topological transition 
to a gapped phase possessing Majorana fermions edge states does not occur. 

\pacs{03.75.Ss, 67.85.Lm, 67.85.-d}
\end{abstract}
\maketitle

%
%

Ultra-cold Fermi atoms are one of the most interesting physical 
systems of the last decade, as they have served as quantum simulators 
of crossover phenomena and phase transitions 
encountered in several areas of physics.
Due to their tunable interactions, 
atoms like $^6$Li and $^{40}$K have been used to study 
the crossover from BCS to BEC superfluidity, 
to simulate superfluidity in neutron stars,
and to investigate unitary interactions which are of great 
interest in nuclear physics.
Furthermore, the ability to control the internal spin state of 
the atoms by using
radio-frequencies (RF) enabled the studies of quantum and classical phase 
transitions as a function of interactions and population imbalance.
These tools have permitted the study of crossover phenomena and
phase transitions, and have validated the symmetry based 
classification of phase transitions put forth by Landau over the
thermodynamic classification proposed earlier by Ehrenfest. 

Very recently a new tool for the toolbox was created:
artificial spin-orbit coupling 
has been produced in neutral bosonic systems~\cite{spielman-2011} where 
the strength of the coupling is controlled optically from weak to strong.
The same technique can be 
applied to ultracold fermions~\cite{spielman-2011} 
and should allow for the exploration of superfluidity 
as a function of interactions and fictitious spin-orbit 
coupling~\cite{chapman-sademelo-2011}.
This possibility created enormous theoretical interest 
recently~\cite{shenoy-2011, chuanwei-2011a,
zhai-2011, hu-2011, iskin-2011, han-2011, chuanwei-2011b}, which was focused 
on zero temperature (ground state) properties. 
Considering possible experiments with 
fermionic atoms such as $^6$Li, $^{40}$K, 
we discuss here topological phase transitions at 
zero and finite temperatures for imbalanced fermions 
during the evolution from BCS to BEC superfluidity 
in three dimensions and in the presence of controllable 
spin-orbit couplings. Even though the symmetry of 
the order parameter does not change through 
topological phase transitions, violating Landau's symmetry-based
classification, clear signatures emerge in 
spectroscopic and thermodynamic properties validating Ehrenfest's
ideas, which combined with 
changes in topological (Hopf) invariants produce a finer
classification scheme of phase transitions.

%
%

{\it Hamiltonian:} We start with the Hamiltonian density
\begin{equation}
{\cal H} ({\bf r})
=
{\cal H}_0 ({\bf r})
+
{\cal H}_I ({\bf r}),
\end{equation}
where the single-particle term is simply
\begin{equation}
\label{eqn:hamiltonian-single-particle} {\cal H}_0 ({\bf r}) =
\sum_{\alpha \beta} \psi^{\dagger}_{\alpha} ({\bf r}) \left[ {\hat
K}_{\alpha} \delta_{\alpha \beta} -h_i ({\bf
r})\sigma_{i,\alpha\beta} \right] \psi_{\beta} ({\bf r}).
\end{equation}
Here, $ {\hat K}_{\alpha} = - \nabla^2/(2 m_{\alpha}) - \mu_{\alpha}
$ is the kinetic energy in reference to the chemical potential
$\mu_{\alpha}$ , and $h_i ({\bf r})$ is the spin-orbit field along
the $i$-direction ($\alpha = \uparrow, \downarrow$, $i=x, y, z$).
The interaction term is
$
{\cal H}_I ({\bf r})
=
-g
\psi^{\dagger}_{\uparrow} ({\bf r})
\psi^{\dagger}_{\downarrow} ({\bf r})
\psi_{\downarrow} ({\bf r})
\psi_{\uparrow} ({\bf r}),
$
where $g$ is a contact interaction, and we set $\hbar =
k_B = 1$.

%
%

{\it Effective Action:} The partition function at temperature $T$ is
$ Z = \int \mathcal{D}[\psi, \psi^\dagger] \exp \left(
 -S[\psi, \psi^\dagger]
\right) $ with action
\begin{equation}
\label{eqn:action-initial}
S[\psi, \psi^\dagger]
=
\int d\tau d {\bf r}
\left[
\sum_{\alpha}
\psi^\dagger_{\alpha} ({\bf r}, \tau)
\frac{\partial}{\partial \tau}
\psi_{\alpha} ({\bf r}, \tau) +
{\cal H} ({\bf r}, \tau)
\right].
\end{equation}

Using the standard Hubbard-Stratanovich transformation that
introduces the pairing field
$
\Delta ({\bf r}, \tau)
=
g
\langle
\psi_{\downarrow} ({\bf r}, \tau)
\psi_{\uparrow} ({\bf r}, \tau)
\rangle
$
and integrating over the fermion variables lead to the effective 
action
$$
S_{\rm eff}
=
\int d\tau d {\bf r}
\left[
\frac{ \vert \Delta ({\bf r, \tau}) \vert^2 }
{g}
-
\frac{T}{2V}
\ln \det \frac{{\bf M}}{T}
+
\widetilde K_+ \delta ({\bf r} - {\bf r}^\prime)
\right],
$$
where
$
\widetilde K_{+}
=
( \widetilde K_\uparrow + \widetilde K_\downarrow )/2.
$
The matrix ${\bf M}$ is
\begin{equation}
\label{eqn:matrix-m}
{\bf M}
=
\left(
\begin{array}{cccc}
\partial_\tau + \widetilde K_\uparrow & - h_\perp & 0 & -\Delta \\
- h_\perp^* & \partial_\tau + \widetilde K_\downarrow &  \Delta & 0 \\
0 & \Delta^\dagger & \partial_\tau - \widetilde K_\uparrow &  h_\perp^* \\
-\Delta^\dagger  & 0 & h_\perp & \partial_\tau - \widetilde K_\downarrow
\end{array}
\right),
\end{equation}
where $h_{\perp} = h_x - i h_y$ corresponds to the transverse component
of the spin-orbit field, $h_z$ to the parallel component
with respect to the quantization axis $z$,
$\widetilde K_\uparrow  = {\hat K}_\uparrow - h_z$,
and $\widetilde K_\downarrow  = {\hat K}_\downarrow + h_z$.

%
%

{\it Saddle Point Approximation:} To proceed, we use the saddle point
approximation $\Delta ({\bf r}, \tau) = \Delta_0 + \eta ({\bf r}, \tau),$ 
and write ${\bf M} = {\bf M}_0 + {\bf M}_F$.
The matrix ${\bf M}_0$ is obtained via the saddle point
$\Delta ({\bf r}, \tau) \to \Delta_0$ which takes ${\bf M} \to {\bf M}_0$,
and the fluctuation matrix ${\bf M}_{{\rm F}} = {\bf M} - {\bf M_0}$
depends only on $\eta ({\bf r}, \tau)$ and its Hermitian conjugate.
Thus, we write the effective action as
$S_{\rm eff} = S_0 + S_{\rm F}$. The first term is 
$$
S_0
=
\frac{V}{T}
\frac{\vert \Delta_0 \vert^2}{g}
-\frac{1}{2}
\sum_{{\bf k}, i\omega_n, j}
\ln
\left[
\frac{i\omega_n - E_j ({\bf k})}{T}
\right]
+
\sum_{\bf k}
\frac{{\widetilde K}_{+}}{T},
$$
in momentum-frequency coordinates $({\bf k}, i\omega_n)$, 
where $\omega_n = (2n + 1) \pi T$. Here, $E_j ({\bf k})$ 
are the eigenvalues of
\begin{equation}
{\bf H}_0
=
\left(
\begin{array}{cccc}
\widetilde K_\uparrow ({\bf k}) & - h_\perp ({\bf k}) & 0 & -\Delta_0 \\
- h_\perp^* ({\bf k}) & \widetilde K_\downarrow ({\bf k}) &  \Delta_0 & 0 \\
0 & \Delta_0^\dagger & -\widetilde K_\uparrow ({-\bf k}) &  h_\perp^* ({-\bf k}) \\
-\Delta_0^\dagger  & 0 & h_\perp (-{\bf k}) & - \widetilde K_\downarrow (-{\bf k})
\end{array}
\right),
\end{equation}
which describes the Hamiltonian of elementary excitations in the
four-dimensional basis $ \Psi^\dagger = \left\{
\psi_{\uparrow}^\dagger ({\bf k}), \psi_{\downarrow}^\dagger ({\bf
k}), \psi_{\uparrow}(-{\bf k}), \psi_{\downarrow}(-{\bf k})
\right\}. $ 
The fluctuation action is 
$$
S_{\rm F}
=
\int d\tau d{\bf r}
\left[
\frac{\vert \eta ({\bf r}, \tau) \vert^2}{g}
-
\frac{T}{2V}
\ln \det
\left(
{\bf 1} + {\bf M}_0^{-1} {\bf M}_{\rm F}
\right)
\right].
$$
The spin-orbit field is $ {\bf h}_\perp ({\bf k}) = {\bf
h}_R ({\bf k}) + {\bf h}_D ({\bf k}), $
where  
$ {\bf h}_R ({\bf k}) = v_R \left( -k_y {\hat{\bf
x}} + k_x {\hat {\bf y}} \right)$ is of 
Rashba-type~\cite{rashba-1984} 
and
$ {\bf h}_D ({\bf k}) = v_D \left( k_y {\hat {\bf
x}} + k_x {\hat {\bf y}} \right) $ 
is of Dresselhaus-type~\cite{dresselhaus-1955},
has magnitude $ h_{\perp} ({\bf k}) = \sqrt{ \left( v_D -
v_R \right)^2 k_y^2 + \left( v_D + v_R \right)^2 k_x^2 }. $ 
For Rashba-only (RO) $(v_D = 0)$ and for equal
Rashba-Dresselhaus (ERD) couplings $(v_R = v_D = v/2)$, the
transverse fields are $ h_{\perp} ({\bf k}) = v_R \sqrt{k_x^2 +
k_y^2}$ ($v_R > 0$) and $ h_{\perp} ({\bf k}) = v \vert k_x \vert $
($v > 0$), respectively.

%
%

{\it Order parameter and number equations:}
The thermodynamic potential is
$\Omega = \Omega_0 + \Omega_F$,
where 
$$
\Omega_0
=
V
\frac{\vert \Delta_0 \vert^2}{g}
-\frac{T}{2}
\sum_{{\bf k}, j}
\ln
\left\{
1 + \exp \left[ - E_j ({\bf k})/T \right]
\right\}
+
\sum_{\bf k}
{\bar K}_{+},
$$
with
$
{\bar K}_{+}
=
\left[
\widetilde K_\uparrow (-{\bf k})
+
\widetilde K_\downarrow (-{\bf k})
\right]/2
$
is the saddle point contribution and 
$\Omega_F = - T \ln Z_F$, with
$
Z_F = 
\int {\cal D} [{\bar \eta}, {\eta}]
\exp 
\left[
- S_F ({\bar \eta}, {\eta}) 
\right]
$
is the fluctuation contribution.
The order parameter is determined via the minimization of $\Omega_0$
with respect to $\vert \Delta_0 \vert^2$ leading to
\begin{equation}
\label{eqn:order-parameter-general} 
\frac{V}{g} 
= 
-\frac{1}{2}
\sum_{{\bf k}, j} 
n_F \left[E_j ({\bf k}) \right] 
\frac{\partial E_j ({\bf k})}{\partial \vert \Delta_0 \vert^2},
\end{equation}
where
$ n_F \left[  E_j (\mathbf{k})  \right] = 1/(\exp\left[ E_j ({\bf
k})/T\right] + 1) $ 
is the Fermi function for energy $E_j ({\bf k})$. We replace the
contact interaction $g$ by the scattering length $a_s$ through the
relation $ 1/g = - m_+/(4\pi a_s) + (1/V) \sum_{\bf k} \left[
1/(2\epsilon_{{\bf k},+}) \right], $ where $ m_+ = 2 m_\downarrow
m_\uparrow / (m_\downarrow + m_\uparrow) $ is twice the reduced
mass, $ \epsilon_{ {\bf k}, \alpha } = k^2 / (2m_\alpha) $ are the
kinetic energies, and 
$ 
\epsilon_{{\bf k}, +} 
= 
\left[ 
\epsilon_{{\bf k}, \uparrow } 
+ 
\epsilon_{ {\bf k}, \downarrow } 
\right] 
/2.
$ 
The number of particles for each spin state is 
$
N_{\alpha} = -
\partial \Omega
/
\partial \mu_{\alpha},
$
which is written as 
\begin{equation}
\label{eqn:number-general}
N_{\alpha} 
= 
N_{\alpha, 0} + N_{\alpha, F},
\end{equation}
where the saddle point contribution is 
$$
N_{\alpha,0}
=
- \frac{\partial \Omega_0}{ \partial \mu_{\alpha} }
=
\frac{1}{2}
\sum_{\bf k}
\left[
1 -
\sum_j n_F \left[ E_j ({\bf k}) \right]
\frac{\partial E_j ({\bf k})}{\partial \mu_{\alpha}}
\right],
$$
and the fluctuation contribution is
$
N_{\alpha, F}
= 
- 
\partial \Omega_F
/
\partial \mu_{\alpha}.
$

The self-consistent relations shown in
Eqs.~(\ref{eqn:order-parameter-general})
and~(\ref{eqn:number-general}) are 
unphysical for atoms of unequal masses, as it is not
possible to have a spin-orbit field 
that converts atom A into atom B, 
as this violates barionic number conservation. 
Therefore, we particularize 
our discussion to balanced and imbalanced systems 
of equal masses, where the explicit eigenvalues of 
the matrix ${\bf H}_0$ are
$
E_1 ({\bf k}) 
=
\sqrt
{
Y_+  + Y_-
+ 
2 \sqrt{
Y_+  Y_-  
- |\Delta_0|^2 |h_\perp|^2
}
},
$
which is always positive definite, and
$
E_2 ({\bf k}) 
=
\sqrt
{
Y_+  + Y_- 
- 
2 
\sqrt
{
Y_+  Y_-  
- |\Delta_0|^2 |h_\perp|^2
}
}
,
$
which can have zeros in momentum space,
$
E_3 ({\bf k}) 
= 
-
E_2 ({\bf k}) 
$
and
$
E_4 ({\bf k}) 
= 
-
E_1 ({\bf k}).
$
Here,
$
Y_+ 
=
{\widetilde K}_+^2 ({\bf k})
+ |\Delta_0|^2 
> 0,
$
$
Y_-
=
{\widetilde K}_-^2 ({\bf k}) 
+ 
|h_\perp ({\bf k})|^2 
> 0,
$
with
$
{\widetilde K}_{\pm} 
= 
(
{\widetilde K}_\uparrow \pm {\widetilde K}_\downarrow
)
/
2.
$
For equal masses, $K_+ = \epsilon_{\bf k} - \mu_+$,
with $\epsilon_{\bf k} = k^2/2m$ being the single
particle dispersion and 
$\mu_+ = ( \mu_\uparrow + \mu_\downarrow )/2$
being the average chemical potential. 
Furthermore, 
$K_- =  ( \mu_\uparrow - \mu_\downarrow )/2 = \mu_- $
plays the role of a Zeeman field that causes population
imbalance $P = N_-/N_+$, where $N_- = N_\uparrow - N_\downarrow$
and $N_+ = N_\uparrow + N_\downarrow$.
We define momentum, energy and velocity scales through the total 
particle density 
$
n_+ 
= 
n_\uparrow 
+ 
n_\downarrow
$ 
or 
$
k_{F+}^3/(3\pi^2) 
= 
k_{F\uparrow}^3/(6\pi^2) 
+ 
k_{F\downarrow}^3/(6\pi^2).
$
This choice leads to the Fermi momentum
$
k_{F+}^3 
=  
(
k_{F\uparrow}^3 
+ 
k_{F\downarrow}^3
)/2,
$
Fermi energy $\epsilon_{F+} = k_{F+}^2/2m$ 
and to the Fermi velocity $v_{F+} = k_{F+}/m$.

\begin{figure} [htb]
\centering
\epsfig{file=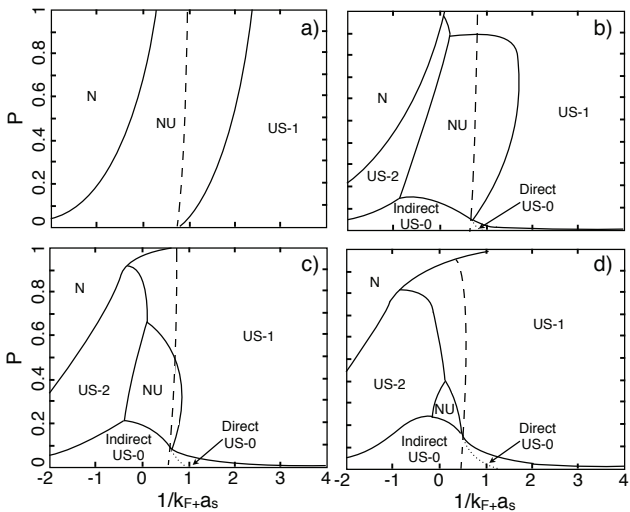,width=1.0 \linewidth} 
\caption{ \label{fig:one} 
Phase diagrams of $P$ versus $1/(k_{F+} a_s)$ for ERD
spin-orbit couplings: 
a) $v/v_{F+} = 0$; 
b) $v/v_{F+} = 0.28$; 
c) $v/v_{F+} = 0.57$; 
d) $v/v_{F+} = 0.85$.
Solid lines are phase boundaries, 
the dashed line corresponds to $\mu_+ = 0$,
and the dotted line corresponds to
a crossover.
}
\end{figure}

{\it Phase Diagram:}
The solutions of Eqs.~(\ref{eqn:order-parameter-general})
and~(\ref{eqn:number-general}) subject 
to thermodynamic stability conditions 
(positiveness of compressibility matrix 
$
\bar \kappa_{\alpha \beta} = 
(\partial N_\alpha/\partial \mu_\beta)_{T,V},
$
isothermal compressibility $\kappa_T$ 
and
volumetric specific heat $C_V$)
lead to the $T = 0$ phase diagram of population imbalance
$P$ versus interaction parameter $1/(k_{F+} a_s)$, which 
is shown in Fig.~\ref{fig:one} for various values of 
ERD spin-orbit coupling $v/v_{F+}$. 
The phase diagram for $v/v_{F+} = 0$ shows a large 
region where a uniform superfluid US phase is unstable, 
we call this region NU for possible non-uniform phases 
including phase separation between superfluid and non-superfluid
components or modulated superfluid phases. 
A general tendency of the phase diagram with increasing 
spin-orbit coupling $v/v_{F+}$ is the stabilization of superfluid
phases and the shrinkage of unstable regions. 
This stabilization is largely due to the emergence
of a triplet component in the order parameter for superfluidity, 
which circumvents the pair breaking abilities of 
Zeeman fields $\mu_-$.  
Many multi-critical points and phases exist, including
a normal (N) and three uniform superfluid regions: 
the phase US-0 corresponds to a fully gapped superfluid with no (zero) 
line of nodes, the phases US-1 and US-2 correspond to 
gapless superfluids with one-line or two-lines of nodes, respectively.
Within the gapped US-0 bulk phase, in our three dimensional
system, a topological transition due to emergence of chiral
edge states (Majorana fermions) does not occur, instead we
have a crossover from indirectly to directly gapped US-0.

%
%

\begin{figure} [htb]
\centering
\epsfig{file=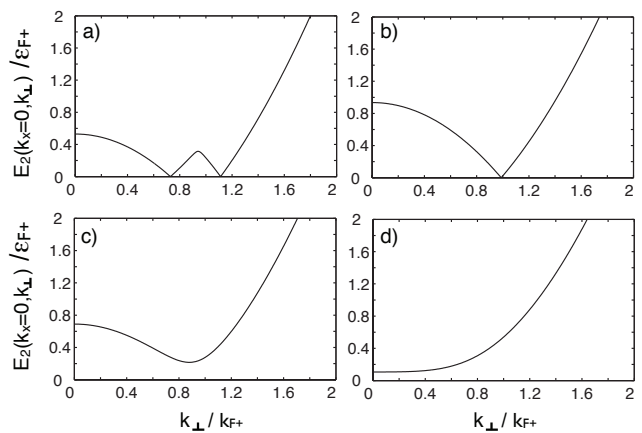,width=1.0 \linewidth}
\caption{ 
\label{fig:two} Excitation energy
$E_2 ({\bf k})/\epsilon_{F+}$ versus $k_\perp/k_{F+}$,
where $k_{\perp} = \sqrt{k_y^2 + k_z^2}$, 
for $v/v_{F+} = 0.57$: 
a) US-2 phase with
$P = 0.5$ and $1/(k_{F+} a_s) = -0.92$; 
b) US-1 phase with
$P = 0.5$ and $1/(k_{F+} a_s) = 2.0$; 
c) US-0 phase (indirect gap) with
$P = 0.06$ and $1/(k_{F+} a_s) = -0.5$; and
d) US-0 phase (direct gap)  with
$P = 0.06$ and $1/(k_{F+} a_s) = 0.5$.
}
\end{figure}

{\it Spectroscopic Properties:}
For the excitation energies $E_1 ({\bf k})$ and $E_2 ({\bf k})$, only
$E_2 ({\bf k})$ can have zeros. In the ERD case,
where $h_\perp ({\bf k}) = v \vert k_x \vert$,
zeros of $E_2 ({\bf k})$ can occur for $Y_+ = Y_-$ 
with $k_x = 0$. 
This leads to the following cases:
(a) two possible lines (rings) of nodes
at 
$
(k_y^2 + k_z^2)/(2m) 
= 
\mu_+ + \sqrt{\mu_-^2 - \vert \Delta_0 \vert^2}
$ 
for the outer ring,
and
$
(k_y^2 + k_z^2)/(2m) 
= 
\mu_+ - \sqrt{\mu_-^2 - \vert \Delta_0 \vert^2}
$ 
for the inner ring,
when $\mu_-^2 - \vert \Delta_0 \vert^2 > 0$;
(b) doubly-degenerate line of nodes 
$
(k_y^2 + k_z^2)/(2m) 
= 
\mu_+ 
$ 
for $\mu_+ > 0$,
doubly-degenerate point nodes
for $\mu_+ = 0$,  or no-line of nodes
for $\mu_+ < 0$,
when 
$\mu_-^2 - \vert \Delta_0 \vert^2 = 0$;
(c) no line of nodes when $\mu_-^2 - \vert \Delta_0 \vert^2 < 0$.
The study of case (a) can be refined into cases
(a2), (a1) and (a0). In case (a2), two rings indeed exist provided that 
$\mu_+ > \sqrt{\mu_-^2 - \vert \Delta_0 \vert^2}$.
However, the inner ring disappears when 
$\mu_+ = \sqrt{\mu_-^2 - \vert \Delta_0 \vert^2}$.
In case (a1), there is only one ring when
$
\vert 
\mu_+ 
\vert 
< 
\sqrt{\mu_-^2 - \vert \Delta_0 \vert^2}.
$
In case (a0), the outer ring disappears at
$\mu_+ = - \sqrt{\mu_-^2 - \vert \Delta_0 \vert^2}$,
and for $\mu_+ < - \sqrt{\mu_-^2 - \vert \Delta_0 \vert^2}$
no rings exist.
Thus, the US-2/US-1 boundary is determined
by the condition $\mu_+ = \sqrt{\mu_-^2 - \vert \Delta_0 \vert^2}$;
the US-0/US-2 boundary is determined by
the Clogston-like condition $\vert \mu_- \vert = \vert \Delta_0 \vert$
when $\mu_+ > 0$,
where the gapped US-0 phase (singlet-rich) 
disappears leading to the gapless US-2 phase (triplet-rich);
and the US-0/US-1 phase boundary is determined by 
$\mu_+ = - \sqrt{\mu_-^2 - \vert \Delta_0 \vert^2}$.
Furthermore, a crossover line between an indirectly gapped 
and a directly gapped 
US-0 phase occurs for $\mu_+ < 0$ and
$\vert \mu_- \vert = \vert \Delta_0 \vert $.
Lastly, some important multi-critical points
arise at the intersections of phase boundaries.
First the point $\mu_+ = 0$ and 
$\vert \mu_- \vert = \vert \Delta_0 \vert$ corresponds
to a tri-critical point for phases US-0, US-1, and US-2.
Second, the point $\vert \Delta_0 \vert = 0$
and $\mu_+  = \vert \mu_- \vert$ corresponds
to a tri-critical point for phases N, US-1 and US-2.
Representative excitation spectra are shown in Fig.~\ref{fig:two}.
Near the zeros of $E_2 ({\bf k})$, quasi-particles have linear
dispersion and behave as Dirac fermions. The change in
nodal structures is associated with bulk topological transitions 
of the Lifshitz class as noted for 
p-wave~\cite{volovik-1992} and d-wave~\cite{duncan-2000} superfluids
The loss of nodal regions correspond to 
annihilation of Dirac quasi-particles with opposite momenta,
which lead to significant changes in spectroscopic properties such as
momentum distributions~\cite{duncan-2000} 
$
n_{{\bf k},\alpha}
=
\left[
1 - 
\sum_{j} n_F [E_j ({\bf k})]
\partial E_j ({\bf k})/\partial \mu_\alpha
\right]
/
2.
$ 
%

%
%

{\it Topological Order:}
The superfluid phases US-0, US-1, US-2 are characterized by 
different effective actions $S_{\rm eff}$ which depend explicitly
on the matrix 
$
{\bf M} (i\omega, {\bf k})
= 
\left[
i\omega {\bf 1} - {\bf H}_0 ({\bf k})
\right]^{-1}.
$
Setting $i\omega \to \omega$ and using algebraic topology~\cite{nakahara-1990},
we construct the topological invariant
$$
m 
=
\int_{\cal D}
\frac{dS_\gamma}{24\pi^2}
\epsilon^{\mu\nu\lambda\gamma}
{\rm Tr}
\left[
{\bf M}\partial_{k_\mu} {\bf M}^{-1}
{\bf M}\partial_{k_\nu} {\bf M}^{-1}
{\bf M}\partial_{k_\lambda} {\bf M}^{-1}
\right]
,
$$
which in the gapped US-0 phase is $m = 0$, 
in the gapless US-1 phase is $m = 1$,
and in the gapless US-2 phase is $m = 2$,
such that $m$ counts the number of rings
in each phase.
The integral above has a hyper-surface measure $dS_\gamma$ 
and a domain ${\cal D}$ 
that encloses the region of zeros of $\omega =  E_j ({\bf k}) = 0$.
Here $\mu, \nu, \lambda, \gamma$ run from 0 to 3,
and $k_\mu$ has components $k_0 = \omega$, 
$k_1 = k_x$, $k_2 = k_y$, and $k_3 = k_z$.

%
%

\begin{figure} [htb]
\centering
\epsfig{file=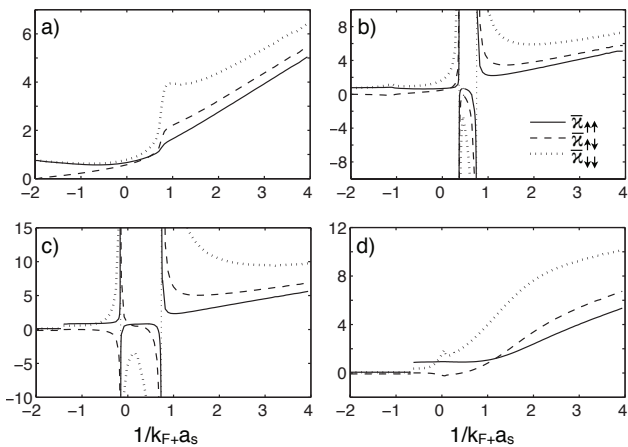,width=1.00 \linewidth} 
\caption{ 
\label{fig:three} The $T = 0$ compressibility matrix 
$\epsilon_{F+}\bar\kappa_{\alpha \beta}/N_+ $ 
as a function of $1/(k_{F+} a_s)$ 
for $v/v_{F+} = 0.57$ at population imbalances
a) $P = 0.06$, b) $P = 0.14$,
c) $P = 0.50$, d) $P = 0.80$. 
}
\end{figure}

{\it Thermodynamic Properties:}
To characterize topological 
phase transitions thermodynamically, 
we show in Fig.~{\ref{fig:three}} 
the compressibility matrix
$
{\bar \kappa}_{\alpha \beta} 
$
as a function of $1/(k_{F+} a_s)$ for $P = 0.06, 0.14, 0.5, 0.8$  
at ERD spin-orbit coupling $v/v_{F+} = 0.57$. 
Notice that singular behavior occurs at phase boundaries.
For instance, in Fig.~{\ref{fig:three}}a, 
${\bar \kappa}_{\alpha \beta}$ has cusps
at the phase boundaries US-2/US-0 and US-0/US-1 and are strictly
positive.
However, in Fig.~{\ref{fig:three}}c, 
${\bar \kappa}_{\alpha \beta}$ has discontinuities
at the N/US-2 boundary, divergences at the US-2/NU and NU/US-1 boundaries
and the diagonal elements 
$\bar \kappa_{\alpha \alpha}$ are negative in the NU region. 
\begin{figure} [htb]
\centering
\epsfig{file=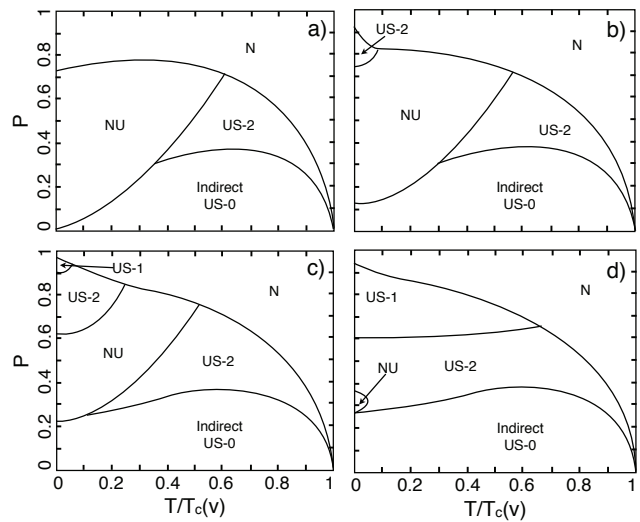,width=1.0 \linewidth} 
\caption{ \label{fig:four} 
Phase diagrams of $P$ versus reduced temperature $T/T_c (v)$
at unitarity $1/(k_{F+} a_s) = 0$ for:
a) $v/v_{F+} = 0$; 
b) $v/v_{F+} = 0.28$; 
c) $v/v_{F+} = 0.57$; 
d) $v/v_{F+} = 0.85$.
}
\end{figure}
%

%
%

{\it Finite Temperatures:}
As experiments are not performed at $T = 0$,
we show in Fig.~\ref{fig:four} the $P$ versus $T$ 
phase diagram at unitarity $1/(k_{F+} a_s) = 0$
for various ERD spin-orbit coupling: a) $v/v_{F+} = 0$;
b) $v/v_{F+} = 0.28$; c) $v/v_{F+} = 0.57$; and d) $v/v_{F+} = 0.85$.
Phase boundaries are determined by 
singular behavior of $\bar \kappa_{\alpha \beta}$. 
In Fig.~\ref{fig:four}, the reduction of
the NU region with increasing $v/v_{F+}$ is due to the increasing
importance of the triplet component of the order parameter, which 
stabilizes phases US-1 and US-2 at intermediate and large $P$.

%
%

{\it Conclusions:} 
We have investigated spectroscopic and thermodynamic properties in 
the BCS-to-BEC evolution at zero and finite temperatures,
including spin-orbit effects.
We have identified bulk topological phase transitions of the Lifshitz 
class between gapped and gapless superfluids and described 
multi-critical points. In addition, we have found that spin-orbit 
effects tend to stabilize uniform superfluid phases against
phase separation 
or non-uniform superfluid phases as population imbalance or 
interactions are changed, due to the creation of a triplet 
component in the order parameter that circumvents pair breaking
effects due to Zeeman fields.

\acknowledgements{We thank ARO (W911NF-09-1-0220) for support.}

\end{document}